# Are Neutrinos Democratic?

G. Karl and J.J. Simpson

Department of Physics, University of Guelph, Guelph, ON N1G 2W1 Canada

**Abstract**

We generalize the notion of democratic mixing matrices for neutrinos and propose a scheme in which the electron neutrino is a superposition of three different mass eigenstates with equal weights. This scheme accounts for the recent SNO results as well as atmospheric $\nu_\mu$ and $\nu_e$ data. The outcomes of reactor neutrino and accelerator experiments are also discussed.

Recently the Sudbury Neutrino Observatory (SNO) has released neutral current data on the total $^8$B solar neutrino flux, in addition to charged-current data on the electron neutrino component [1,2]. The data show that, to a good approximation, one-third of the initial $^8$B neutrinos arrive at the detector with their original flavor ($\nu_e$), while two-thirds arrive with a modified flavor ($\nu_\mu/\nu_\tau$). Together with atmospheric neutrino data [3,4] there is now compelling evidence for neutrino oscillations. A number of mixing schemes are being actively pursued, such as bimaximal mixing [5], tri-bimaximal mixing [6] and democratic mixing [7]. As there are precisely three neutrino flavors, the SNO result encourages us to speculate on a mixing matrix where the oscillation of $\nu_e$ to one-third each of $\nu_e$, $\nu_\mu$ and $\nu_\tau$ is the exact result at large distances. We find that a simple model can account reasonably well for SNO, Superkamiokande (SuperK) and several other experiments.

A simple way to ensure that an incident $\nu_e$ beam behaves at large distances as one third $\nu_e$, and two thirds $\nu_\mu$, and $\nu_\tau$ is to assume that it oscillates into one third each, which can be achieved if the electron neutrino is an equal superposition of three mass eigenstates $\nu_1$, $\nu_2$, and $\nu_3$ with different masses. The other two weak-flavor states are linear superpositions orthogonal on the electron neutrino.

The simplest such mixing matrix is a permutation of the "democratic" mixing matrix [7]:

$$\begin{pmatrix} \nu_e \\ \nu_\mu \\ \nu_\tau \end{pmatrix} = \begin{pmatrix} 1/\sqrt{3} & 1/\sqrt{3} & 1/\sqrt{3} \\ 0 & 1/\sqrt{2} & -1/\sqrt{2} \\ -2/\sqrt{6} & 1/\sqrt{6} & 1/\sqrt{6} \end{pmatrix} \begin{pmatrix} \nu_1 \\ \nu_2 \\ \nu_3 \end{pmatrix} \quad (1)$$

This mixing matrix differs from the democratic matrices in the literature by identifying the first row with the neutrino of the lightest charged lepton, $\nu_e$. It is easy to check that with this matrix $\nu_e$ evolves, at large distances, into an equal mixture of the three different weak-flavor neutrinos. However, the matrix (1) is only the simplest example of a democratic matrix which has $\nu_e$ as an equal superposition of the three mass eigenstates. More generally one can leave the first row of the matrix fixed and rotate the second and third row in their common plane by an angle $\alpha$. Then the mixing matrix will depend on the angle $\alpha$ and have the form:

$$\begin{pmatrix} \nu_e \\ \nu_\mu \\ \nu_\tau \end{pmatrix} = \begin{pmatrix} 1/\sqrt{3} & 1/\sqrt{3} & 1/\sqrt{3} \\ -\frac{2\sin\alpha}{\sqrt{6}} ; & \frac{\sin\alpha}{\sqrt{6}} + \frac{\cos\alpha}{\sqrt{2}} ; & \frac{\sin\alpha}{\sqrt{6}} - \frac{\cos\alpha}{\sqrt{2}} \\ -\frac{2\cos\alpha}{\sqrt{6}} ; & \frac{\cos\alpha}{\sqrt{6}} - \frac{\sin\alpha}{\sqrt{2}} ; & \frac{\cos\alpha}{\sqrt{6}} + \frac{\sin\alpha}{\sqrt{2}} \end{pmatrix} \begin{pmatrix} \nu_1 \\ \nu_2 \\ \nu_3 \end{pmatrix} \qquad (2)$$

Note that when $\alpha = 0$ the matrix (2) becomes the same as the matrix (1). It can be checked that the mixing matrix (2) is orthogonal, and one can compute that <u>at large distances</u> (classical limit), the probabilities $P_{ee} = P_{e\mu} = P_{e\tau} = 1/3$ and $P_{\mu\mu} = P_{\tau\tau} = 1/2$ while $P_{\mu\tau} = 1/6$ (for any $\alpha$). The full formulae for $P_{ee}$, $P_{\mu\mu}$, and $P_{e\mu}$ will be given further on.

This mixing scheme would seem to be ruled out by the results of Kamiokande [3] and SuperK [4] which usually are taken to indicate that atmospheric electron neutrinos do not oscillate in traveling from the upper atmosphere to the detector. However, the data only show that with an initial flux of two parts $\nu_\mu$ and one part $\nu_e$, the $\nu_\mu$-component diminishes while the $\nu_e$-component does not. It turns out that the democratic mixing matrix above mimics this behaviour. In the classical limit, we obtain the following:

muon neutrino flux:     Initial $\langle\nu_\mu\rangle = 2$ ;     Final $\langle\nu_\mu\rangle = 2 \times P_{\mu\mu} + 1 \times P_{\mu e} = 4/3$ electron neutrino flux:     Initial $\langle\nu_e\rangle = 1$ ;     Final $\langle\nu_e\rangle = P_{ee} + 2 \times P_{e\mu} = 1$.

The $v_e$-flux remains constant by being replenished by $v_\mu$ oscillations. The ratio of ratios R can also be computed:

$$R = \frac{(<v_\mu>/<v_e>)observed}{(<v_\mu>/<v_e>)expected} = \frac{(4/3/1)}{(2/1)} = \frac{2}{3} = 0.667$$

to be compared to the values quoted by SuperK [4], which are:

$R_{sub\ GeV} = 0.652 \pm 0.019$ (stat) $\pm 0.051$ (syst)

$R_{multi\ GeV} = 0.661 \pm 0.034$ (stat) $\pm 0.079$ (syst)

Clearly, the classical probabilities are in agreement both with the solar neutrino fluxes and the ratio R of SuperK, provided neutrino masses allow the classical limit to be obtained. So it is worth investigating this scheme in more detail.

Under the mixing matrix (2), the evolution probability $P_{ee}$ is given by

$$P_{ee} = P(v_e \to v_e) = \frac{1}{3}\left(1 + \frac{2}{3}(\cos\phi_{12} + \cos\phi_{32} + \cos\phi_{13})\right) \to \frac{1}{3} \qquad (3)$$

where the phase angles are

$$\phi_{ij} = \frac{(m_i^2 - m_j^2)L}{2E} = \frac{\Delta m_{ij}^2 L}{2E}, i,j = 1,2,3. \qquad (4)$$

The $m_i$ are neutrino masses, L the distance from the source of neutrinos, and E the neutrino energy. Also

$$\Delta m_{32}^2 = \Delta m_{12}^2 - \Delta m_{13}^2 \qquad (5)$$

One sees that, in the classical limit where the cosine terms average to zero, $P(v_e \to v_e) = \frac{1}{3}$, which is in agreement with the SNO results [1,2] and with SuperK's solar neutrino results [8,2]. One only needs to choose neutrino masses such that the classical limit is obtained and matter effects in the sun are insignificant, and this can be achieved with the same $\Delta m_{ij}^2$ that are appropriate for atmospheric neutrino oscillations discussed below.

The formulae for $P_{\mu\mu}$ and $P_{e\mu}$ are much more lengthy and are given in an appendix. Their classical limits are as noted above: $P_{e\mu} \to 1/3$ and $P_{\mu\mu} \to 1/2$.

It can be checked that the probabilities $P_{ee} + P_{e\mu} + P_{e\tau} = P_{\mu\mu} + P_{e\mu} + P_{\mu\tau} = P_{\tau\tau} + P_{e\tau} + P_{\mu\tau} = 1$, that the diagonal probabilities $P_{ee} = P_{\mu\mu} = P_{\tau\tau} = 1$ at distances $L = 0$, that the off-diagonal probabilities vanish at $L = 0$, and that the classical values are obtained as quoted before.

The distribution of L/E for surviving atmospheric $\nu_\mu$ is shown in Fig. 1. Because the reactor experiments CHOOZ [9] and Palo Verde [10] rule out $\Delta m^2 \geq 10^{-3} eV^2$, the distribution for $\Delta m_{12}^2 = 0.7 \times 10^{-3}$ and $\Delta m_{13}^2 = 0.9 \times 10^{-3}$ and $\alpha = \pi/4$ is shown. Although no attempt has been made to fit the data shown [4] to determine the best set of parameters, $\alpha = \pi/4$ gives considerably better agreement than $\alpha = 0$, and is perhaps marginally better than $\alpha = \pi/2$. The case of $\alpha = \pi/4$ is interesting, as it implies an equal mixture of $\nu_\mu$ and $\nu_\tau$ as the preferred basis states for neutrino oscillations. The L/E distribution for surviving atmospheric $\nu_e$ is practically constant, in agreement with the data of ref. 4.

The present scheme makes some predictions for the outcomes of new and imminent experiments. KamLAND is an experiment to detect $\overline{\nu_e}$ from reactors within 150-200 km distance of a scintillator detector in the Kamiokande mine in Japan [11]. The phase factors are very large for the masses appropriate for atmospheric neutrinos, and hence there will be rapid fluctuations of survival probability as a function of $\overline{\nu_e}$-energy which will be averaged to a flux one-third of the expected flux due to energy binning and variation in reactor distances.

The K2K experiment [12] is an experiment to look for $\nu_\mu$ disappearance of 1.3 GeV neutrinos over a baseline of 250 km, $P_{\mu\mu}$. If the $\Delta m^2$ and $\alpha$ are approximately equal to the values we used for atmospheric neutrinos, K2K will see about 97% of the full flux of $\nu_\mu$ for $\alpha = \pi/4$ or $\pi/2$. There would be a small contribution of $\nu_e$ at the detector, about 1% for $\alpha = \pi/4$ and 3% for $\alpha = \pi/2$. An observation by K2K of substantial muon neutrino disappearance would be strong evidence against the present scheme, if all $\Delta m^2$ must be less than $10^{-3}$ eV$^2$.

Finally we should remark that SNO may soon make a more accurate determination of the total flux from the neutral-current reaction on deuterium, and it will be interesting to see if the 1/3 ratio of $\nu_e$ to $\nu_{total}$ is maintained, since the present model requires this to be exact.

We are aware that this scheme conflicts with the gallium measurements of solar $\nu_e$ flux [13-15]. However, we assign more weight to the SNO and Superkamiokande solar neutrino results because these experiments determine both the $\nu_e$ and the total $\nu$ flux.

Democratic matrices studied in the literature [7] make a different identification of the electron neutrino, constrained by detailed assumptions about lepton mass matrices. We do not attempt to understand neutrino masses, focusing entirely on the oscillation phenomena. This is the source of the departure of the mass matrix (1) from the literature.

In conclusion, with the assumption that the electron neutrino is an exact one-third mixture of three massive neutrinos, chosen to account for the SNO charged-current/neutral current ratio, we have found that we can account for atmospheric $\nu_\mu/\nu_e$ ratio of ratios essentially exactly. We can account for the fact that the atmospheric $\nu_e$ flux is not diminished nor shows a zenith-angle dependence, and that the atmospheric $\nu_\mu$-flux does show a strong zenith angle dependence. In addition, we suggest that the K2K experiment may be crucial to testing this mixing scheme. These results are all achieved with $\Delta m_{ij}^2$ values appropriate to atmospheric oscillations, of the order $10^{-3}$ eV$^2$.

**Appendix:** Formulae for $P_{ij} = P\nu_i\nu_j(\alpha)$

Equation (3) in the text gives $P_{ee}$ which does not depend on $\alpha$. We give below $P_{e\mu}$ and $P_{\mu\mu}$ which depend on the angle $\alpha$ and are lengthier:

$$P_{e\mu} = \left(\frac{1}{9}\right)(\cos\phi_{12} + \cos\phi_{13})(2\cos^2\alpha - 2) + (\sqrt{3}/9)(\cos\phi_{13} - \cos\phi_{12})\sin 2\alpha$$
$$+ \left(\frac{1}{9}\right)(\cos\phi_{32})(1 - 4\cos^2\alpha) + 1/3$$

$$P_{\mu\mu} = \left(\frac{1}{2}\right) + \left(\frac{2}{9}\right)[(\cos\phi_{12} + \cos\phi_{13})(1 + \cos^2\alpha - 2\cos^4\alpha)$$
$$+ \sqrt{3}(\cos\phi_{13} - \cos\phi_{12})(\sin 2\alpha)(\cos^2\alpha - 1)]$$
$$+ \left(\frac{1}{18}\right)(\cos\phi_{32})(1 - 8\cos^2\alpha + 16\cos^4\alpha)$$

From these two equations one can derive the other probabilities $P_{e\tau}$, $P_{\mu\tau}$ and $P_{\tau\tau}$ by probability conservation, but these are not needed in the computations reported above. It can be checked that for $\phi_{ij} = 0$, $P_{e\mu} = 0$ and $P_{\mu\mu} = 1$, while if we average over $\phi_{ij}$, $P_{e\mu} = 1/3$ and $P_{\mu\mu} = 1/2$.

**References**

[1]    Q.R. Ahmad et al., nucl-ex/0204008

[2]    Q.R. Ahmad et al., Phys. Rev. Lett. 87 (2001) 071301

[3]    Y. Fukuda et al., Phys. Lett. B335 (1994) 237

[4]    H. Sobel, Proc. of the XIXth International Conference Neutrino Physics and
       Astrophysics. Nucl. Phys. B (Proc. Suppl.) 91 (2001) 127

**Fig. 1** Atmospheric $\nu_\mu$ survival probability as a function of the L/E ratio. The squares are the data presented in Fig. 7 of [4]. The solid line shows the averaged results of the present theory with $\Delta m_{12}^2 = 0.7 \times 10^{-3}$ eV$^2$, $\Delta m_{13}^2 = 0.9 \times 10^{-3}$ and $\alpha = \pi/4$. Data and theory are normalized at the two smallest L/E values.

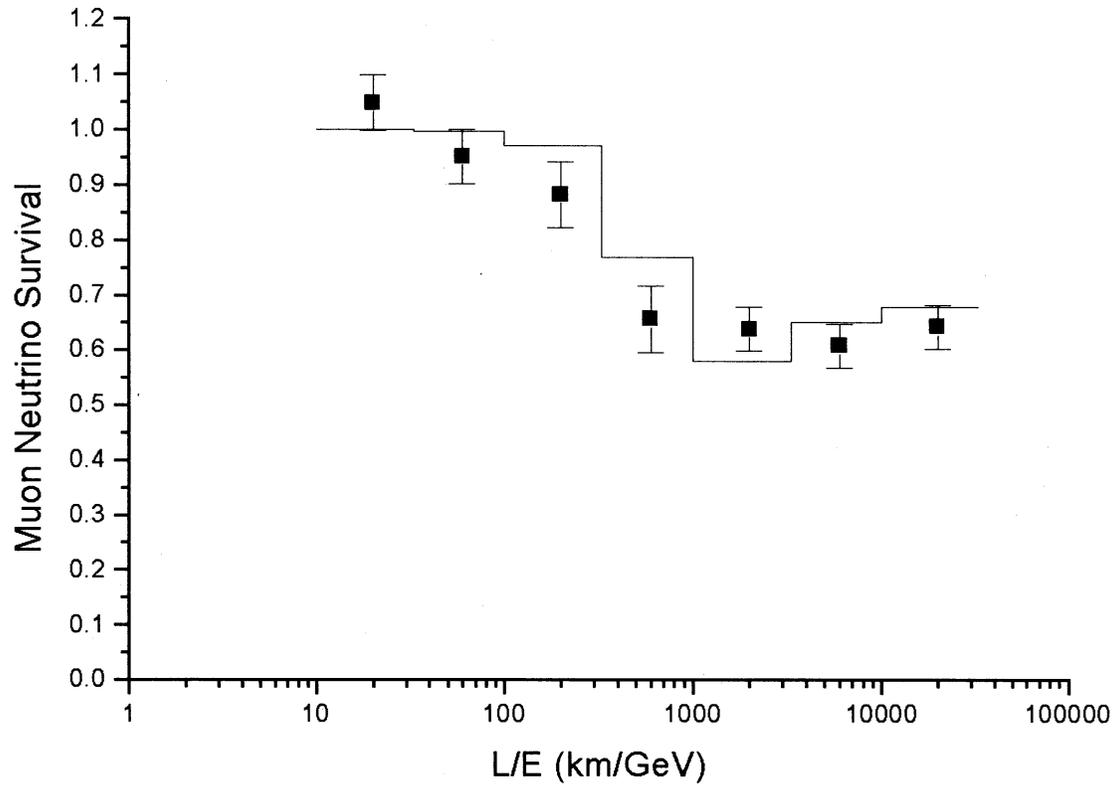